\documentclass[]{aastex}
\usepackage{emulateapj5}
\usepackage[]{graphicx}

\newcommand{\kms}{km~s$^{-1}$}
\newcommand{\ha}{H$\alpha$}
\newcommand{\hi}{\ion{H}{1}}
\newcommand{\hii}{\ion{H}{2}}
\newcommand{\nii}{[\ion{N}{2}]}
\newcommand{\vi}{\hbox{$V\!-\!I$}}
\newcommand{\emax}{\hbox{$e_{max}$}}

\slugcomment{Accepted by The Astronomical Journal}

\shorttitle{Double Bars, Inner Disks, and Nuclear Rings}
\shortauthors{Erwin \& Sparke}

\begin{document}

\title{Double Bars, Inner Disks, and Nuclear Rings in Early-Type Disk
Galaxies}

\author{Peter Erwin}
\affil{Instituto de Astrofisica de Canarias, C/ Via L\'{a}ctea s/n, 
38200 La Laguna, Tenerife, Spain}
\email{erwin@ll.iac.es}
\and
\author{Linda S. Sparke}
\affil{University of Wisconsin-Madison, 475 North Charter Street, 
Madison, WI 53706}
\email{sparke@astro.wisc.edu}

\begin{abstract}
  
We present results from a survey of an unbiased sample of thirty-eight
early-type (S0--Sa), low-inclination, optically barred galaxies in the
field, using images both from the ground and from space.  Our goal was
to find and characterize central stellar and gaseous structures:
secondary bars, inner disks, and nuclear rings.  We find that bars
inside bars are surprisingly common: at least one quarter of the
sample galaxies (possibly as many as 40\%) are double-barred, with no
preference for Hubble type or the strength of the primary bar.  A
typical secondary bar is $\sim12$\% of the size of its primary bar and
extends to 240--750 pc in radius.  Secondary bars are not
systematically either parallel or perpendicular to the primary; we see
cases where they lead the primary bar in rotation and others where
they trail, which supports the hypothesis that the two bars of a
double-bar system rotate independently.  We see \textit{no}
significant effect of secondary bars on nuclear activity: our
double-barred galaxies are no more likely to harbor a Seyfert or LINER
nucleus than our single-barred galaxies.

We find kiloparsec-scale inner disks in at least 20\% of our sample;
they occur almost exclusively in S0 galaxies.  These disks are on
average 20\% the size of their host bar, and show a wider range of
relative sizes than do secondary bars.  Nuclear rings are present in
about a third of our sample.  Most of these rings are dusty, sites of
current or recent star formation, or both; such rings are
preferentially found in Sa galaxies.  Three S0 galaxies (8\% of the
sample, but 15\% of the S0's) appear to have purely stellar nuclear
rings, with no evidence for dust or recent star formation.

The fact that these central stellar structures are so common indicates
that the inner regions of early-type barred galaxies typically contain
dynamically cool and disklike structures.  This is especially true for
S0 galaxies, where secondary bars, inner disks, and/or stellar nuclear
rings are present at least two thirds of the time.  If we interpret
nuclear rings, secondary bars, and (possibly) inner disks and nuclear
spirals as signs of inner Lindblad resonances (ILRs), then between one
and two thirds of barred S0--Sa galaxies show evidence for ILRs.

\end{abstract}


\section{Introduction}

Isolated examples of multiply barred galaxies --- where one or more
small (``secondary'') bars reside concentrically within a larger bar
--- have been known for over twenty-five years, the first such cases
having been pointed out by \citet{deV74}.  However, in the last decade
high-resolution imaging has revealed numerous double-barred galaxies;
see \citet{friedli96b} for an early review.  Interest in multiply
barred galaxies was stimulated by the suggestion of \citet{shlosman89}
that independently rotating, concentric bars within bars could be an
important mechanism for feeding gas into the centers of galaxies,
potentially fuelling nuclear activity.  Subsequent n-body +
hydrodynamic simulations \citep{friedli93b,combes94} showed how
independently rotating secondary bars might form within large-scale
bars, and \citet{witold00} showed that such systems could be
dynamically self-consistent.  Models in which the secondary bars
rotate at the same speed as the primary bars have been proposed by
\citet{shaw93} and \citet{heller96}, though the lack of preferred
relative orientations in known double-barred galaxies seems to support
the idea of independent rotation \citep{buta93,friedli93b,friedli96b}.

Double- and even triple-barred galaxies continue to turn up in a
variety of imaging surveys
\citep{w95,mrk,jungwiert97,martini99,marquez99,greusard00,laine02}. 
However, we still do not know how common such structures are in
ordinary galaxies.  None of the studies cited used a complete,
unbiased sample; the majority were specifically focused on Seyfert
galaxies and ``matching'' control galaxies.  In addition, other
substructures within large-scale bars may be confused with secondary
bars.  \citet{seifert96}, in a study of edge-on S0 galaxies,
identified several examples of inner disks, distinct from both the
bulge and the outer disk, and \citet{vdb98} found an edge-on S0 galaxy
with both a nuclear disk and a stellar nuclear ring just outside it. 
\citet{erwin99} showed that stellar nuclear rings and inner disks
inside the large-scale bars of moderately inclined galaxies could be
mistaken for secondary bars --- or even coexist with them.  Failure to
distinguish between these structures can yield false detection rates
for double bars and obscure the true diversity of structures inside
bars.

In this paper we present results from a survey of a complete sample of
early-type, optically barred galaxies, using the 3.5m WIYN telescope
and archival optical and near-IR images from the \textit{Hubble Space
Telescope} (HST)\footnote{The WIYN Observatory is a joint facility of
the University of Wisconsin-Madison, Indiana University, Yale
University, and the National Optical Astronomy Observatories. 
Observations with the NASA/ESA Hubble Space Telescope obtained at the
Space Telescope Science Institute, operated by the Association of
Universities for Research in Astronomy, Inc., under NASA contract
NAS5-26555.}, intended to find and characterize multiple bars and
other central features.  Erwin \& Sparke (2002, hereafter Paper~II)
provides the details for the individual galaxies.  Here, we discuss
the types of structures that we found --- both stellar and gaseous
---, their frequencies, and their relation to other properties of the
galaxies, including nuclear activity.

\section{Sample Selection and Observations}

Details of our sample and observations are presented in Paper~II; here
we give the basic outlines.  We selected all barred S0--Sa galaxies in
the UGC catalog \citep{ugc} north of $-10\arcdeg$ in declination, with
heliocentric radial velocity $\leq 2000$ \kms, major axis diameter
$D_{25} \ge 2\arcmin$, and ratio of major to minor axis $a/b \leq 2$,
corresponding to inclinations $\lesssim 60\arcdeg$.  Galaxies in the
Virgo Cluster were excluded, since there is evidence that Hubble types
for Virgo galaxies disagree with those for field galaxies
\citep[][]{koopman98}.  The final sample had a total of 38 galaxies,
listed in Table~\ref{tab:galaxies}: twenty S0's, ten S0/a's, and eight
Sa's (twenty-five of the galaxies are SB, with the remainder SAB). 
The median distance\footnote{We assume $H_{0} = 75$ \kms{} Mpc$^{-1}$. 
Distances to about one third of the galaxies are from
surface-brightness fluctuation measurements; see Paper~II.} to
galaxies in the sample is 18.5 Mpc.

Because our galaxies are selected on the basis of being
\textit{optically} barred --- that is, SB or SAB according to RC3 ---
it is possible that we missed some barred galaxies mistakenly
classified as unbarred.  Conflicting claims have been made about
whether and how much the bar fraction increases when galaxies are
observed in the near-infrared, where dust extinction is less of a
problem; the most extensive survey to date is that of
\citet{eskridge00}, who classified 186 disk galaxies using
\textit{H}-band images.  They found that the primary difference in
going from RC3 optical classifications to the near-IR was the increase
in the relative number of strong (SB) bars; the \textit{total} bar
fraction (SB + SAB) increases by less than 10\%.  Not surprisingly,
this effect is weakest for early type galaxies: 65\% of S0--Sab
galaxies in their sample are barred according to the RC3, while 71\%
are barred in the near-IR. Since we restrict ourselves to S0--Sa
galaxies, and select both SB and SAB classes, we are probably missing
only one or two barred galaxies which were mistakenly classified as
SA.

All but two of the galaxies were observed in $B$ and $R$ with the
3.5-m WIYN Telescope, under generally excellent seeing conditions
(median seeing of 0.8\arcsec{} in $R$, corresponding to about 70 pc at
the median distance).  In addition, we found HST archival images ---
WFPC2 and/or NICMOS --- for just over half the sample.  The most
common filters used were F606W and F814W with WFPC2 and F160W with
NICMOS.

\section{Analysis and Classifications\label{sec:analysis}}

\begin{figure}
\begin{center}
\includegraphics[scale=0.92]{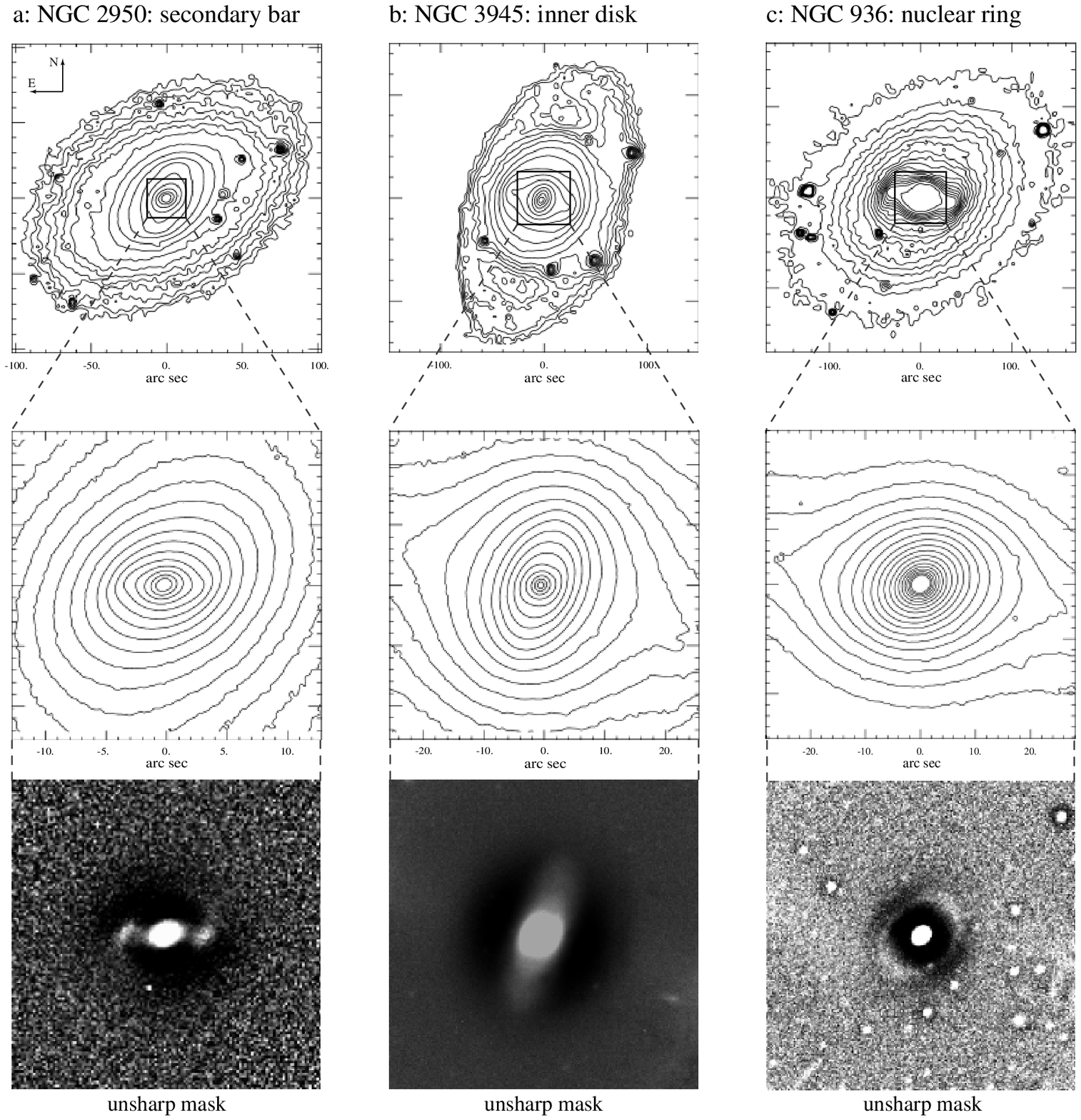}
\end{center}

\caption{Examples of stellar structures inside the bars of three SB0
galaxies.  For each galaxy, we show $R$-band isophotes (top and
middle, except that the top part of \textbf{c} is a DSS image) and
unsharp masks made from the $R$-band images (bottom).  \textbf{a.)}
Inside the primary bar of NGC 2950 is a secondary bar; the
double-lobed structure in the unsharp mask is characteristic of bars. 
\textbf{b.)} Inside the primary bar of NGC 3945 is an inner disk ---
an elliptical feature at the same position angle as the galaxy's outer
disk.  \textbf{c.)} Inside the bar of NGC 936 is a stellar nuclear
ring.  As in NGC 3945, the elliptical feature is aligned with the
galaxy's outer disk, but here the unsharp mask indicates a ring. 
\textit{All of these structures have the same signature in ellipse
fits: a peak in ellipticity at fixed position angle}.  (HST images
indicate that NGC 2950 also has a weak, stellar nuclear ring
surrounding the secondary bar, and that NGC 3945 has both a stellar
nuclear ring \textit{and} a small secondary bar deep inside its inner
disk; see Erwin \& Sparke 1999 and Paper~II.)\label{fig:demo-stellar}}

\end{figure}

\begin{figure}
\begin{center}
\includegraphics[scale=0.92]{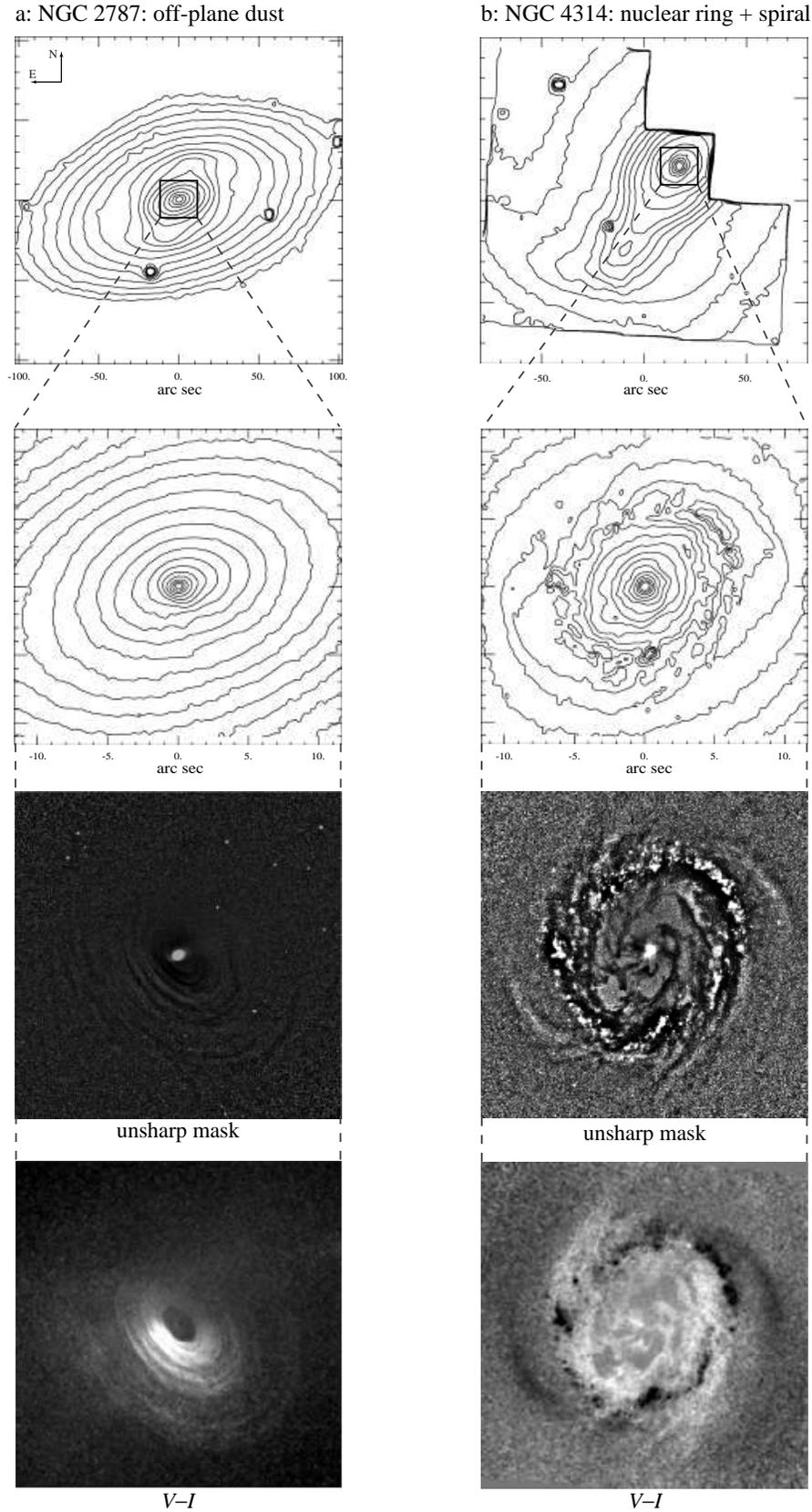}
\end{center}

\caption{Examples of gaseous structures inside bars.  We show, from
top to bottom: $R$-band (top left only) and WFPC2 F814W isophotes;
unsharp masks; and WFPC2 \vi{} color maps (light = red, dark = blue). 
\textbf{a.)} Inside the bar of the SB0 galaxy NGC 2787 is a
spectacular tilted dust disk; \hi{} outside the optical disk is also
misaligned with respect to the stars \citep[][]{shostak87}. 
\textbf{b.)} Inside the primary bar of the SBa galaxy NGC 4314 is a
star-forming nuclear ring, with both dust lanes (red) and sites of
recent star formation (blue); further inside is a dusty nuclear
spiral.  (We also find an inner disk outside the off-plane dust in NGC
2787, and HST near-IR images confirm a previously identified secondary
bar inside NGC 4314's nuclear ring; see Paper~II for
details.)\label{fig:demo-gas}}

\end{figure}

We identify two general classes of structures inside bars: stellar and
gaseous.  The first class includes \textit{secondary bars} (also
called \textit{inner bars}), \textit{inner disks}, and \textit{stellar
nuclear rings}; see Figure~\ref{fig:demo-stellar} and \citet{erwin99}
for examples.  The second class includes \textit{dusty and
star-forming nuclear rings}, \textit{nuclear spirals}, and
\textit{off-plane dust structures} (small polar rings and inclined
dust disks); see Figure~\ref{fig:demo-gas}.  Because our
classifications are based on broad-band optical and near-IR images, we
use the presence of dust and star formation (as indicated by color
maps) as a proxy for the presence of gas.

A detailed discussion of our methods can be found in Paper~II. To
summarize briefly: we noted candidate features from ellipse fits to
the $R$-band isophotes (for HST images, we use the reddest available
filter, typically F814W for WFPC2 and F160W for NICMOS).  We focused
on both local peaks \textit{and} troughs in ellipticity profiles, as
well as significant deviations in position angle for the fitted
ellipses \citep[see][for an example of a secondary bar detected in
NGC~3945 as a \textit{minimum} in ellipticity, due to projection
effects]{erwin99}.  The nature of individual ellipse-fit features was
then checked by inspecting the images, by unsharp masking, and by the
use of color maps.  This let us discriminate between bars, rings, and
spirals (e.g., Figure~\ref{fig:demo-stellar}) --- all of which can
produce similar ellipse-fit features --- and determine when dust
extinction might be responsible.  In addition, unsharp masking and
color maps revealed features not immediately apparent in the ellipse
fits, including some nuclear rings; color maps are a standard way of
identifying and measuring dusty and star-forming nuclear rings
\citep{buta93}.  For off-plane dust identification, we also required
kinematic evidence from the literature of misalignment between stars
and gas in the galaxy.

Although it may be possible to distinguish between inner bars and
disks using unsharp masks (cf.\ Figure~\ref{fig:demo-stellar}), we use
a conservative classification based on orientation: an elliptical
stellar structure aligned within ten degrees of the outer disk is
classified as an inner disk, unless unsharp masking shows that it is
clearly a ring --- compare Figure~\ref{fig:demo-stellar}b and c. 
(While we also required that an inner disk's apparent ellipticity be
less than that of the outer disk, we found no cases of \textit{more}
elliptical structures aligned with the outer disk.)  Thus, we are
likely to classify some inner bars with chance alignments --- and
possibly some poorly resolved stellar rings as well --- as inner
disks; the category may also include highly flattened inner bulges. 
However, in Section~\ref{sec:inner-disks} we argue that inner disks
differ statistically from inner bars, and form a genuinely distinct
class.

Stellar structures such as bars, disks, and (stellar) rings can be
obscured by dust.  Consequently, we include the category ``dusty'' for
galaxies where the central regions (roughly, $0.5\arcsec \lesssim r
\lesssim 10\arcsec$) are too confused by dust and/or star formation
for us to determine if there is an inner \textit{stellar} structure. 
This is rarely a problem where HST NICMOS images are available. 
Intrinsically dusty structures such as nuclear spirals and dusty
nuclear rings are much easier to find; we are likely to miss these
only if they are too small to be properly resolved.  Extremely strong
\ha{} + \nii{} emission could contaminate some $R$-band images;
however, this usually occurs together with star formation and
accompanying dust, situations where we rely on NICMOS images or else
assign a ``dusty'' classification.

Table~\ref{tab:galaxies} lists the structures detected in each galaxy;
Paper~II provides measurements and analysis of the individual
galaxies.

\section{Demographics of Inner Structures}\label{sec:demo}

In Table~\ref{tab:freq} we list the distribution of central
structures, sorted by Hubble type and bar strength.  In each row, we
give the fraction of all galaxies, followed by the fraction of each
Hubble type or bar type, which have the specified structure; stellar
structures are listed first (also including ``dusty'' galaxies, where
we were unable to identify central stellar structures), followed by
gaseous structures; as noted above, the latter are generally based on
detection of dust and/or star-formation.  The final section of the
table groups structures by their relevance to whether or not a galaxy
has an inner Lindblad resonance (see below).  The most striking
results are the unexpectedly high frequency of both secondary bars and
inner disks, and the discordance of their distribution among Hubble
types.

At least a quarter of the galaxies are multiply barred (first row of
Table~\ref{tab:freq}).  If we assume that the ``dusty'' galaxies are
no more or less likely to harbor secondary bars than the rest of the
sample, and exclude them from the statistics, then the frequency of
secondary bars may be as high as 40\%.  There is no obvious trend with
Hubble type: Sa galaxies are as likely as S0's to be double-barred.

The fraction of galaxies with inner disks is also quite high: we find
them in at least $\sim20$\% of the sample, or 28\% of the unobscured
galaxies.  Since it is easier, using our techniques, to detect inner
\textit{bars} in near--face-on galaxies than it is to find inner
disks, the inner-disk frequency could be as high as that of secondary
bars.  (Conversely, a secondary bar that is aligned with the outer
disk, or an unresolved stellar nuclear ring, could be classified in
error as an inner disk: at the highest level of unsharp masking,
NGC~4386's inner disk bears some resemblance to a bar.)  In contrast
to the inner bars, inner disks show a marked preference for earlier
Hubble types: all but one of the disks are found in S0's, and none at
all in Sa's.  The single S0/a galaxy with an inner disk is NGC~4643,
and its inner disk shows some signs of being a nuclear ring rather
than a disk (Paper~II).  In Section~\ref{sec:inner-disks} we consider
the question of whether inner disks are genuinely distinct from inner
bars.

Nine galaxies (24\% of the sample, or 36\% of the unobscured galaxies)
have neither an inner disk nor a secondary bar\footnote{A careful
reader may notice that the percentages of inner bars, inner disks, and
the lack thereof do not sum to 100\% for the unobscured galaxies. 
This is because NGC~3945 has both an inner bar \textit{and} an inner
disk.}.  NGC~3032, NGC~4203, NGC~4665, NGC~5701, and NGC~7743 have no
central structures other than (in some cases) dusty nuclear spirals,
while NGC~936, NGC~2273, NGC~4245 and NGC~5377 have nuclear rings
and/or spirals but no inner bars or disks.

At least eleven galaxies (29\% of the sample) have nuclear rings. 
Eight of these rings show up in color maps as either red, blue, or a
mixture of both: they are dusty and/or sites of current or recent star
formation.  But in NGC~936, NGC~2950, and NGC~3945, the rings appear
purely stellar and are the same color as the surrounding stars; these
were identified from ellipse fits and unsharp masking.  The
distribution of dusty and star-forming nuclear rings is clearly skewed
towards the later Hubble types: the frequency is 10\% for S0's versus
50\% for the Sa galaxies.  This is not surprising, given that Sa
galaxies have more gas in the disk than S0 galaxies
\citep[e.g.,][]{roberts94}.

Nine galaxies show evidence for dusty nuclear spirals (24\% of the
sample).  The nuclear spiral in NGC~2273 includes two arms which are
both blue and IR-bright, which indicates that these are star-forming
sites, and the same may be true in NGC~3032.  In the rest, we see no
evidence for enhanced star formation, and the spirals appear to be
dust structures only.  As for dusty/star-forming nuclear rings, 
nuclear spirals are more common in later Hubble types, though the 
trend is not as strong.

All of the features discussed above can, with some degree of
confidence, be linked with the presence of inner Lindblad resonances
(ILRs) within bars.  The case is strongest for nuclear rings, which
are generally understood as the result of one or two ILRs acting on
bar-driven gas flows.  Most theoretical models of double-bar systems
\citep[][]{pfenniger90,friedli93b,shaw93,witold00} also link the
presence and dynamics of a secondary bar to the primary bar's ILR(s). 
The case for inner disks is less clear; however, some could be
mis-classified secondary bars or nuclear rings, and stars on the
$x_{2}$ orbits associated with ILRs might also produce an inner disk,
so we include them as well.  Finally, nuclear spirals have also been
linked to the influence of ILRs on bar-driven gas flows
\citep[e.g.,][]{englmaier00,witold02}, though acoustic spirals might
not depend on ILRs as directly \citep{elmegreen98}.

We list the possible ILR-signifiers in decreasing order of confidence
in the bottom section of Table~\ref{tab:freq}.  If only nuclear rings
are considered, then 29\% of galaxies have signatures of ILRs; if
secondary bars, inner disks, and nuclear spirals are also counted,
this fraction becomes 66\%.  Thus, at least one third of our galaxies
show good evidence for ILRs, and the fraction could be two thirds or
even higher (since we may have missed detecting some secondary bars,
inner disks, and nuclear rings).  These signatures appear to be most
common in the S0 galaxies.  The apparent deficiency in S0/a galaxies
is not statistically significant.

\textit{None} of these inner structures shows any particular
preference for bar strength.  Within the statistical uncertainties,
they are all --- secondary bars, inner disks, nuclear rings, and
nuclear spirals --- as likely to be found in SB galaxies as in SAB
galaxies.

Finally, there are six galaxies with good evidence for off-plane gas
in the nuclear regions, based both on our images and kinematic
evidence from the literature.  Three of these were previously known or
suspected (NGC~2685, NGC~4203, and NGC~7280; see discussions in
Paper~II for references).  All but one of these candidate polar-ring
galaxies are S0's --- in fact, they constitute 25\% of the S0's.

\subsection{Characteristics of Secondary Bars\label{sec:inner-bars}}

\begin{figure}
\begin{center}
\includegraphics[scale=0.45]{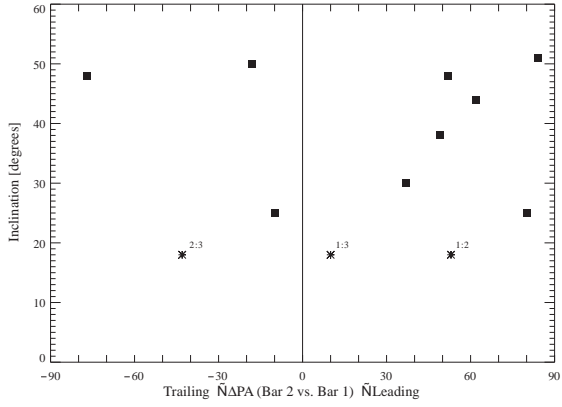}
\end{center}

\caption{Relative position angles of secondary and primary bars,
corrected for projection effects.  The stars indicate the three bars
of NGC 2681; the adjacent numbers indicate which of its three bars are
being compared (1 = primary, 2 = secondary, 3 =
tertiary).\label{fig:db-leadtrail}}

\end{figure}

To investigate the orientation and sizes and of secondary bars, we
deprojected their measured position angles and sizes, assuming that
the bars are flat, linear structures (see Paper~II for position angles
and inclinations used).  Figure~\ref{fig:db-leadtrail} shows the
relative position angles of bars in the double-barred galaxies.  We
used the same scheme as \citet{buta93} and \nocite{w95}Wozniak et al.\
(1995; hereafter W95): secondary bars can ``lead'' or ``trail'' the
primary; we determine the sense of rotation from the observed spiral
structure in the disk, by assuming that the spirals are trailing
(Table~5 of Paper~II).  (The implicit assumption behind this scheme is
that both bars are rotating, and that they rotate in the same
direction as the trailing spiral pattern.)  Counting all the bar
pairs, there are eight cases where the inner bar leads the outer, and
four cases where it trails.  The excess of leading bars is not
statistically significant; the presence of both leading and trailing
secondary bars agrees with the results of Buta \& Crocker, W95, and
\citet{friedli96a}, and supports the idea that the two bars rotate
independently of one another.

\begin{figure}
\begin{center}
\includegraphics[scale=0.45]{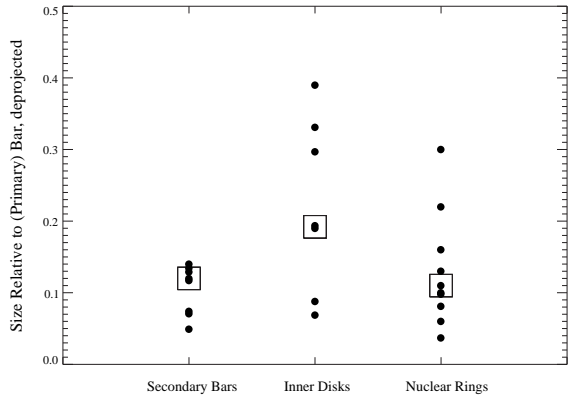}
\end{center}

\caption{The sizes of secondary bars, inner disks, and nuclear rings,
as a fraction of the host bar's length (the host bar is the galaxy's
primary bar --- or only bar, if it has no secondary bar).  The open
squares indicate the median relative size for each type of structure
(excluding both inner bars of NGC 2681, for reasons given in the
text).\label{fig:relsize}}

\end{figure}

Secondary bars have a fairly limited range of sizes relative to the
primary bar: they have lengths which are 0.05--0.14 times the lengths
of their primary bars, with a median of 0.12
(Figure~\ref{fig:relsize}).  Here, length refers to the deprojected
upper limit on bar semi-major axis, as given in Paper~II; we excluded
NGC~2681, since it is unclear which of its three bars should be
considered the ``secondary.''  These figures are close to, but
generally smaller than, those of \nocite{w95}W95: their double bars
had a median relative size of 0.15, and a range of 0.06--0.27, if we
exclude NGC~2681 and the three ``B+B+B'' galaxies in their sample. 
\citep[These numbers do not change appreciably if we instead use the
near-IR measurements from][]{friedli96a}.  Since the W95 sample was
chosen on the basis of previously suspected inner features --- mostly
from photographic images --- and was observed with lower resolution
than our galaxies, there was almost certainly a selection effect in
favor of large inner bars.

\begin{figure}
\begin{center}
\includegraphics[scale=0.45]{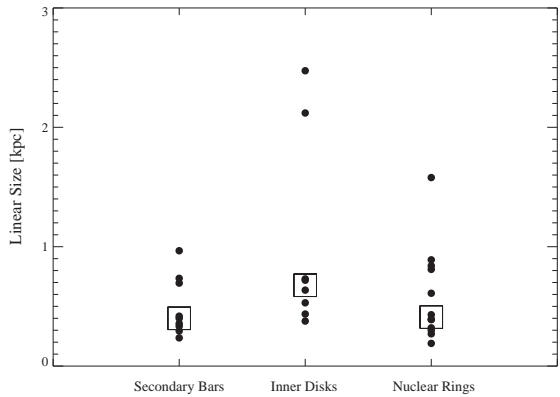}
\end{center}

\caption{As for Figure~\ref{fig:relsize}, but here giving the linear
sizes, in kpc, of the various inner structures.  The open squares
indicate the median size for each type of structure (again, excluding
NGC 2681).\label{fig:kpcsize}}

\end{figure}

In linear terms, secondary bars have a median semi-major axis $a
\approx 400$ pc, with the smallest being 240 pc and the largest about
1 kpc; see Figure~\ref{fig:kpcsize}.  This corresponds to a range of
3\% to 8\% (median 5\%) of the parent galaxy's $R_{25}$ (half of the
$\mu_{B} = 25$ diameter from \nocite{rc3}RC3).  This is, not
surprisingly, much smaller than the $R_{25}$-fractional sizes of the
large-scale bars in our sample (14--100\%); however, some single bars
in Sbc and later spirals can be this small \citep{martin95}.

\begin{figure}
\begin{center}
\includegraphics[scale=0.45]{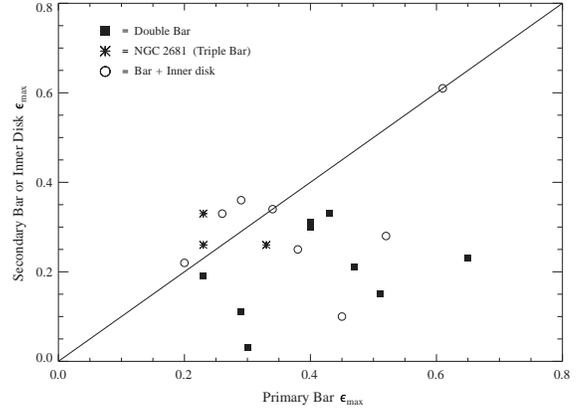}
\end{center}

\caption{Measured isophotal (i.e., not deprojected) ellipticity of
secondary bars (squares), inner disks (circles), and the multiple bars
of NGC 2681 (stars) as a function of primary bar
ellipticity.\label{fig:emax}}

\end{figure}

\begin{figure}
\begin{center}
\includegraphics[scale=0.45]{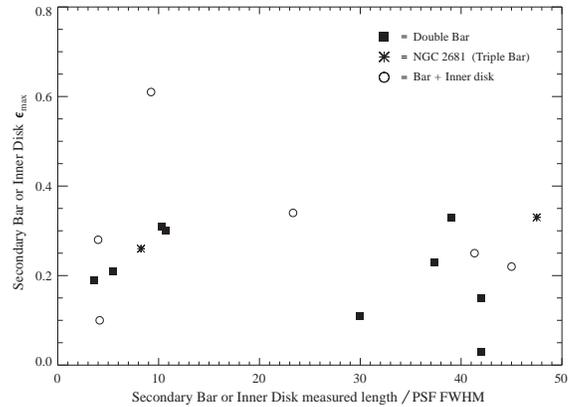}
\end{center}

\caption{Measured isophotal ellipticity of secondary bars (secondary
and tertiary bars for NGC 2681) and inner disks as a function of
secondary bar or disk size, relative to the resolution of the image
used to measure ellipticity (resolution defined as full-width
half-maximum of point-spread function).  Points farther to the right
are better resolved.\label{fig:emax-v-length}}

\end{figure}

Figure~\ref{fig:emax} displays the maximum isophotal ellipticity
measured for secondary bars versus the ellipticity of their primary
bars.  Secondary bars appear systematically rounder than primary bars,
as found by \nocite{w95}W95.  The secondary bars might be
intrinsically rounder, or the isophotes could be rounder due to the
superposition of light from the bright inner regions of round bulges;
circularization of the isophotes due to seeing effects is another
possibility.  Disentangling the first two possibilities will probably
require complex decomposition of the images, but we can judge the
effect of seeing more simply.  In Figure~\ref{fig:emax-v-length}, we
compare the observed ellipticity $\emax$ for secondary bars and inner
disks with the observed (not deprojected) length of the bar or disk
divided by the full-width half-maximum (FWHM) of the image's
point-spread function (PSF).  The isophotal ellipticity of secondary
bars and inner disks appears independent of resolution: features which
are small compared with the observational resolution are not
systematically rounder than large features --- in fact, the secondary
bars with the lowest isophotal ellipticity are among the best
resolved!  We conclude that the observed lower ellipticity of
secondary bars reflects mostly a combination of intrinsic roundness
and the superposition of bulge light; it is not due to seeing effects
\nocite{laine02}(Laine et al.\ 2002 make a similar argument).

The triple-barred galaxy NGC~2681 is a bit of a puzzle.  The
middle-sized secondary bar is \textit{more} elliptical than the
largest bar, which is atypical, and the length ratio of these
two is 0.31 --- much higher than that of any of the double bars.  If
we instead regard the tertiary (innermost) and secondary as a
bar-within-bar system, we find a length ratio of 0.17 --- larger than
any of the double bars, but not dramatically so --- and the tertiary
is less elliptical than the secondary.  The secondary-tertiary system
thus appears similar to ``normal'' double-barred galaxies, while the
primary-secondary system is anomalous.

Do secondary bars have distinct colors?  In four cases (NGC~2681,
NGC~2859, NGC~2962, and NGC~4314) we can identify distinct, asymmetric
color features (always red) aligned or associated with the secondary. 
In NGC~2681, NGC~2962, and NGC~4314, the large size of the secondary
bar (in NGC~2681) or HST images let us see that the color variations
are clearly due to individual dust lanes, which in NGC~4314 partly
resemble the classic leading-edge dust lanes found in large-scale bars
and in simulations of gas flow in barred galaxies (see
Figure~\ref{fig:demo-gas}; \nocite{martini01}Martini et al.\ 2001
noted the existence of similar ``leading-edge'' dust lanes in the
inner bars of Mrk 270 and Mrk 573).  The dust lanes in NGC~4314 could
also be interpreted as a nuclear spiral, which is the conservative
classification we use in Table~\ref{tab:galaxies} and Paper~II. We
suspect that the red, asymmetric features aligned with the secondary
bar in NGC~2859 are probably dust as well, though we lack the
resolution to be sure.  In four more galaxies (NGC~2950, NGC~3945,
NGC~6654, and NGC~7280), there is little or no sign of color features
associated with the secondary bars, other than a symmetric inward
reddening trend which follows the isophotes (NGC~2950 and NGC~6654);
our resolution is not high enough to tell if the innermost red regions
in NGC~718 and NGC~3941 are actually aligned with their secondary
bars.  There is no sign in any of these galaxies that the inner bars
are composed of young, blue stars.  We conclude that secondary bars
are probably made up of stars not too different from those of the
primary bar and bulge.

We also note that several secondary bars appear relatively free of
in-plane, corotating gas.  In particular, NGC~2950, NGC~3945, and
NGC~6654 have only a few, isolated patches of dust (or none at all),
while in NGC~3941 and NGC~7280 the dust lanes and gas kinematics
suggest that the gas is off-plane.

The relative sizes of these observed double-bar systems suggests that
the double bars created in n-body + gas simulations are not yet good
models of real double bars: the simulated inner bars are simply too
large.  The two inner bars formed in \nocite{friedli93b}Friedli \&
Martinet's (1993) simulations were 0.26 and 0.5 times the size of the
outer bars, respectively; the relative size of the inner bar in
\citet{friedli96a} was 0.21.  On the other hand, the double-bar system
in Model~IV of \citet{rautiainen99} --- who performed n-body
simulations with \textit{no} gas component --- has a relative
size\footnote{Measured at $t = 12.5$ Gyr from the amplitude spectra in
their Figure~8, using the outermost radius of the 0.06 contours, and
from the second and fifth contour levels in the particle density plots
of Figure~11.} of $\sim 0.15$, which is on the upper end of the range
we find.

\subsection{Inner Disks}\label{sec:inner-disks}

Are the structures that we identify as inner disks physically distinct
from inner bars?  Our definition (Section~\ref{sec:analysis}) perforce
includes secondary bars which happen to have the same orientation as
the outer disk, unless they are sufficiently bright and narrow to
produce isophotes more elliptical than those of the outer disk. 
Nonetheless, there are three reasons why we believe that our inner
disks form a distinct population of structures.

First, we find too many inner disks for all of them to be chance
alignments of inner bars.  We classify inner elliptical features as
inner disks if the observed (projected) position angle is within
10\arcdeg{} of the galaxy's line-of-nodes.  At the median inclination
of our galaxies (48\arcdeg), this corresponds to a putative bar lying
within roughly $15\arcdeg$ of the line of nodes.  By chance, then, we
should expect an average of 17\% of inner bars to be aligned closely
enough with the outer disk for us to mis-classify them as inner disks. 
This would imply about two misclassified inner bars in addition to the
ten that we do identify --- instead, we see \textit{eight} inner
disks.

Second, inner disks show a wider range of relative sizes than do inner
bars (Figure~\ref{fig:relsize}).  They vary from $\sim 7$\% of primary
bar length to almost 40\%, with a median size of 19\%; the latter is
well outside the range spanned by the inner bars (0.05--0.14).  A
Kolmogorov-Smirnov test comparing the relative sizes gives a
probability of 96\% that the two distributions are drawn from
different parent populations.  (The linear and $R_{25}$-fractional
sizes of inner disks have more overlap with those of inner bars,
though the upper bounds for inner disks are clearly higher: $a = 260$
pc to 1.7 kpc, with a median of 470 pc, or 3\% to 22\% of $R_{25}$,
with a median of 7\%; see Figure~\ref{fig:kpcsize})

Third, there is the strong difference in frequency with Hubble types. 
All but one of the inner disks are found in S0 galaxies, while inner
bars occur with equal frequency in the S0/a and Sa galaxies of our
sample, and have been noted previously in galaxies as late as Sbc 
\citep[e.g.,][]{jungwiert97}.

Most of the inner disks are roughly the same color as the surrounding
bar or bulge.  In NGC~4143 and NGC~4643 there is evidence for nuclear
spirals in the color maps, but the disks themselves are not distinct
color features.  The inner disks of NGC~2685 and NGC~4386, however,
\textit{are} quite distinct, and are redder than their surroundings. 
Both disks appear to have axisymmetric color profiles; our resolution
is not good enough to tell whether this represents a gradient in
stellar population, or dust obscuration.

There is one case of a secondary bar residing inside an inner disk
(NGC~3945); this is also the only clear example of an inner disk
coexisting with a nuclear ring.  We found no cases of inner disks
inside secondary bars, though these would be small structures and 
might have remained unresolved in our images.

\subsection{Nuclear Rings}\label{sec:nr}

Nuclear rings may be more common than our detection rate of 29\%, for
a couple of reasons.  Dusty and star-forming nuclear rings are easier
to find than stellar rings, since, rather than being obscured by dust,
they are often part of the dust obscuration.  But a sufficiently small
nuclear ring can still escape detection, at least in our ground-based
images.  Indeed, if we consider only those galaxies with archival HST
images (just over half the sample), then the nuclear ring fraction
rises to 38\%.  Finding \textit{stellar} nuclear rings also requires
an absence of dust \textit{and} high signal-to-noise data (as is the
case for NGC~936; see Paper~II).  Thus, we suspect that there are
probably more stellar nuclear rings in our sample than the three we
identify.

There are three galaxies with star-forming nuclear rings in our
survey: NGC~2273, NGC~4245, and NGC~4314; the last is a well-studied
prototype of this class of nuclear ring.  Taken together, they
constitute $17 \pm 9$\% of the S0/a and Sa galaxies, a figure roughly
consistent with the estimated 10\% of Sc and earlier spirals thought
to have circumnuclear star-forming rings, based on an HST UV imaging
survey \citep{maoz96,maoz01}.  Three more galaxies have smooth, blue
nuclear rings: NGC~718, NGC~2681, and the inner of NGC~5377's two
nuclear rings.  These may represent later stages in the star-formation
process, where active star formation has mostly ceased and clumped
distributions have begun to smooth out.  If we lump the star-forming
and blue nuclear rings together, then it appears that one-third of the
S0/a--Sa galaxies have experienced circumnuclear starbursts in the
last Gyr or so.

No S0 galaxies have star-forming or blue nuclear rings, and only one
has a dusty nuclear ring.  On the other hand, all three of the stellar
nuclear rings are in S0 galaxies.  This suggests that circumnuclear
star formation \textit{did} take place in at least some S0 galaxies,
probably several Gyr ago.  An alternate hypothesis is that the stellar
rings in NGC~2950 and NGC~3945 are dynamical side-effects of their
double-bar natures.  \citet{masset97} showed that rings could form in
n-body simulations due to the nonlinear coupling between two different
pattern speeds (bar and spiral, in their simulations); such stellar
rings might also form in double-barred systems if the corotation of
the small bar matched the ILR of the large bar (M. Tagger, private
communication).  This cannot explain \textit{all} the stellar rings,
however, since NGC~936 shows no signs of an inner bar.  Similarly,
there is no sign of an inner bar in the Virgo SB0 NGC~4371, which has
a prominent stellar nuclear ring \citep{erwin99}.

The nuclear rings we find have a range of sizes similar to, but
slightly larger than, those of the inner bars
(Figures~\ref{fig:relsize} and \ref{fig:kpcsize}).  The median size
relative to their parent bars is 11\%, with a range of 3.7--30\% (the
upper limit is set by the large blue ring surrounding the middle bar
of NGC~2681; without it, the maximum size is 22\%).  The linear size
range is $a = 190$--1500 parsecs, with a median of 410 parsecs; the
$R_{25}$-fractional sizes are 2.3--17\% (median = 5.3\% of $R_{25}$). 
All these measurements are quite consistent with those of the nuclear
rings in \nocite{buta93}Buta \& Crocker's (1993) compilation.  The
median semi-major axis of their nuclear rings is 5.4\% of $R_{25}$,
with a range of 1.2--26\% (using $D_{25}$ values from
\nocite{rc3}RC3).  They do not give bar sizes for their galaxies; but
if we use the inner-ring sizes in their compilation as estimates of
the bar sizes \citep[since inner rings are typically only slightly
larger than the bars they surround;][]{buta86,buta95}, we can derive
bar-relative sizes for their nuclear rings.  This gives relative sizes
of 3.7--32\%, with a median of 11\% --- essentially identical to what
we find for our nuclear rings.

Simulations of gas flow in barred galaxies \citep[e.g.,][]{athan92,
shaw93, piner95} suggest that nuclear gas rings can lie both parallel
and perpendicular to their host bars, or else at oblique angles with
the major axis of the ring leading the bar \citep[but
see][]{heller01}.  Is this true for our galaxies?  Of the eleven rings
in our sample, six appear to be leading, one is apparently
perpendicular to its bar, and three are approximately circular.  Only
NGC~4314 has an elliptical ring which trails its (primary) bar
slightly, something rarely if ever seen in any single-bar simulations. 
This may be due to the fact that NGC~4314 is double-barred, something
we discuss below.

\subsubsection{Secondary Bars and Nuclear Rings}\label{sec:nr-db}

As first noted by \citet{buta93} and \citet{shaw93}, nuclear rings can
coexist with secondary bars.  In fact, this is rather common in our
sample: six of the ten double-barred galaxies (NGC~718, NGC~2681,
NGC~2859, NGC~2950, NGC~3945, and NGC~4314) have nuclear rings
surrounding the secondary bars.  Comparison with single-barred
galaxies (excluding the dusty galaxies) shows that double-barred
galaxies are far more likely to also have nuclear rings: only 27\% of
the single-barred galaxies have nuclear rings, versus 60\% of the
double-barred galaxies.  On the other hand, secondary bars do not
\textit{require} nuclear rings: NGC~2692, NGC~3941, NGC~6654, and
NGC~7280 all feature secondary bars without accompanying rings.

In five galaxies, we can make reasonably accurate measurements of the
relative sizes of nuclear rings and secondary bars.  The secondary bar
in NGC~3945 is $\sim 65$--100\% of the size of the surrounding nuclear
ring (the range comes from comparing the measured length of the bar to
both minor and major axes of the ring); in NGC~2681 and NGC~2950, the
secondary bars appear to terminate in the rings; the secondaries in
NGC~2859 and NGC~4314 are $\sim$ 90\% and 80\% the size of their
respective nuclear rings.  Since nuclear rings are generally believed
to mark the approximate location of a bar's inner Lindblad
resonance(s) \citep[e.g.,][]{athan92,buta93,piner95}, this suggests
that the inner bars extend no further than their primary bars' ILRs. 
\nocite{pfenniger90}Pfenniger \& Norman's (1990) suggestion was that a
secondary bar's corotation should coincide with the primary's inner
Lindblad resonance.  This situation seems to hold in the simulations
of \citet{friedli93b} and \citet{combes94}, and \citet{witold00} were
able to construct a dynamically plausible double-bar model using a
particular version of that condition.  The observed sizes of secondary
bars and nuclear rings are at least consistent with this ILR-CR
hypothesis.  The overall range of absolute and relative sizes reported
above for \textit{all} our secondary bars and nuclear rings also
agrees with this: secondary bars as a class fall into the same size
range as nuclear rings, though the upper limit on secondary bar size
appears to be lower than that of nuclear rings.

We can also use the relative orientations of nuclear rings and
secondary bars to test alternate theories of double-barred galaxies,
particularly those where secondary bars are side effects of nuclear
rings and corotate with the primary bar.  There are no examples in our
sample of \textit{parallel} inner and outer bars separated by a
nuclear ring, which is one plausible scenario \citep[cf.\ NGC~4321;
see][and references therein]{knapen00}.  However, \citet{shaw93} and
\citet{heller96} both produced models where a nuclear ring could
induce a corotating secondary bar misaligned with the primary.  The
hydrodynamical + N-body simulations of Shaw et al.\ produced a
secondary bar parallel with the elliptical, gaseous nuclear ring and
leading the primary bar.  Heller \& Shlosman, on the other hand, found
that a massive, elliptical nuclear ring (not necessarily gaseous)
which led the primary bar could cause the $x_{1}$ orbits interior to
the ring to twist as much as 20\arcdeg{} in the \textit{trailing}
direction.  The resulting secondary bar would then trail the primary
bar and would thus \textit{not} be aligned with the nuclear ring.

However, neither picture can explain the configurations of all of our
secondary bar + nuclear ring systems.  NGC~2859 \textit{is} consistent
with the model of \citet{shaw93} --- the ring is parallel with the
secondary, and both lead the primary bar by $\approx 85\arcdeg$ (after
deprojection); but in NGC~4314, the secondary bar and the ring are
parallel but \textit{trail} the primary by $\approx 10\arcdeg$.  The
mechanism of \citet{heller96} produces a trailing secondary bar, but
it requires that an elliptical nuclear ring lead the primary bar by a
significant amount (60\arcdeg, in their example).  It thus predicts
that the secondary bar should be strongly \textit{mis}aligned with the
ring, which is clearly \textit{not} true for NGC~4314.  In contrast,
the simulations of \citet{friedli93b} produce independently rotating
secondary bars with aligned, gaseous nuclear rings; this appears to be
the best explanation for a system like NGC~4314 \citep[cf.\ the
hydrodynamical simulations of this galaxy by][who found that a
faster-rotating inner bar produced better agreement with the observed
dust lanes than a corotating inner bar]{ann01}.  In NGC~2681 and
NGC~2950, approximately \textit{circular} nuclear rings encloses the
secondary bars, which both trail (in NGC~2681) and lead (in NGC~2950)
the primary bars.  Again, such configurations do not agree with the
corotating double-bar models.

\subsection{Nuclear Activity}

Of the 38 galaxies in our sample, we were able to find spectral
classifications of their nuclei in the literature for 31; these are
summarized in Table~\ref{tab:agn}.  Following \citet{ho97b}, we
consider galaxies with Seyfert, LINER, and transition (``T'') nuclei
to be AGNs --- that is, the emission-line spectra in those galaxies
result at least partly from something other than star formation and
associated \hii{} regions.  Those galaxies with \hii{} or ``SB''
(starburst) classifications are lumped together as \hii{} nuclei.

To compare the nuclear-activity distributions with the results of
\nocite{ho97b}Ho et al., we consider galaxies in the S0 and S0/a--Sab
lines from their Table~2A. Combining these, we find $63\pm4$\% AGNs,
$15\pm3$\% \hii{} nuclei, and $22\pm3$\% absorption-line nuclei, which
is similar to what we see for our sample (first line of our
Table~\ref{tab:agn}).  This is also consistent with their finding
(\nocite{ho97c}Ho, Filippenko, \& Sargent 1997b) that the frequencies
of nuclear activity were similar for both barred and unbarred
early-type galaxies.

Secondary bars do \textit{not} appear to enhance --- \textit{or}
reduce --- nuclear activity significantly.  Neither is there a
significant difference when we consider galaxies with axisymmetric
centers (inner disks or ``none'' category).  Galaxies with secondary
bars or inner disks may tend to lack \hii{} nuclei, but the numbers 
involved are small.

Nuclear activity \textit{is} related to the presence of a
dusty/star-forming nuclear ring, a nuclear spiral, or off-plane dust
near the nucleus.  \textit{None} of the galaxies with these features
has an absorption-line nucleus.  This is consistent with the findings
of \citet{martini99}: 20 of the 24 Seyfert galaxies that they studied
with HST had nuclear spirals, while only 5 had nuclear-scale bars.

Bar \textit{strength} may also be a factor: the SAB galaxies in our
sample are more likely than the SB galaxies to host AGNs.  While this
is on the borderline of being statistically significant (at
approximately the 2.5-sigma level), it does agree with the conclusions
of \citet{shlosman00}, who found, using several different samples and
several measures of bar strength, that AGNs were found in galaxies
with weak bars more often than in those with strong bars.

\section{Discussion}

\subsection{Comparison with Previous Studies}

As we argued in the introduction, there have been no studies directly
comparable to ours, since previous studies have looked at small
samples, heterogeneous or biased samples, or focused on specific types
of galaxies (such as Seyferts).  Nonetheless, we would like to know if
our findings --- particularly the high frequencies of inner bars and
disks --- agree, in general, with what has been found before.

\citet{mulchaey97} examined matched samples of Seyfert and control
galaxies in the $K$ band \citep{mrk}, covering a range of Hubble types
from S0 to Sc.  They reported that 20\% of their control galaxies were
double-barred, but only 10\% of the Seyferts.  (Considering only
galaxies with at least a large-scale bar, the inner-bar fractions are
$9 \pm 6$\% and $22 \pm 10$\%, respectively.)  These figures are
clearly lower than our double-bar detection rate.  The discrepancy
becomes worse if some of their inner bars are misidentified inner
disks or nuclear rings (e.g., NGC~2273; see Paper~II).  Using
kinematic distances from LEDA, we find a median distance of 34.5 Mpc
for their galaxies; combined with their median seeing of 1.0\arcsec,
this yields a typical linear resolution of 170 pc, compared with 70 pc
for our ground-based images.  We suspect that a number of inner bars
in their galaxies may simply have been missed due to the lower
resolution, a possibility they noted.  Of course, there is also a
difference in the samples: their galaxies were either Seyferts, or
specifically chosen to match the Seyferts, and covered a wider range
of Hubble types (S0--Sc).

The $H$-band imaging survey of \citet{jungwiert97} includes many more
nearby galaxies; its only drawback is that it is apparently a subset
of a larger, unfinished sample, with no selection criteria specified. 
Among their galaxies, a total of 35 barred systems meet our primary
selection criteria (axis ratio $< 2$ and redshift $< 2000$ \kms),
although they cover a much wider range in Hubble type (S0--Sd).  The
typical linear resolution was similar to that of our ground-based
images: their median seeing of 1.0\arcsec{}, combined with a median
galaxy distance of 19 Mpc, gives a resolution of 90 pc.  Although they
did not have HST data, their use of $H$-band images does mean they
were usually less affected by dust obscuration than we are.  Eleven of
these galaxies are identified by Jungwiert et al.\ as having secondary
bars, for a double-bar frequency of $31 \pm 8$\%, in good agreement
with what we find.  Since Jungwiert et al.\ did not attempt to
distinguish inner bars from inner disks, their double-bar frequency is
in fact significantly \textit{lower} than our combined inner-bar plus
inner-disk frequency (45\% of our whole sample, 64\% of our unobscured
galaxies).  This supports our finding that the frequency of inner
disks declines abruptly for Hubble types later than S0: only two of
their eleven double-barred galaxies are S0.

More recently, \citet{laine02} used a combination of archival HST
NICMOS $H$-band images and Digitized Sky Survey images to examine
matched sets of Seyfert and ``control'' galaxies, similar in many
respects to those of \citet{mrk}.  The median distance of their
galaxies (31 Mpc) was quite similar, but the use of HST data means
their resolution is much higher; consequently, it is not surprising
that they report a higher rate of secondary bars than did
\citet{mulchaey97}: $28 \pm 5$\% of their barred galaxies have inner
bars.  As with the preceding studies, there was no attempt to
distinguish inner bars from inner disks or nuclear rings, and their
strict ellipse-fit criteria almost certainly means that some bars or
disks were missed (e.g., NGC~4143 and NGC~7280; see Paper~II). 
Nonetheless, the agreement with our results is good, given the broader
range in Hubble type (S0--Sc; only three of their twenty multiply
barred galaxies were S0's).

\citet{seifert96} looked for small-scale disks inside edge-on S0
galaxies, and found them in approximately half their sample (7 or 8
out of 15 galaxies).  Since their galaxies were at roughly the same
distances as ours, and since the inner disks they found have similar
sizes\footnote{Measured from peaks in their $a_{4}/a$ plots, which are
the most appropriate match to our radii of maximum ellipticity.},
these are likely to be edge-on analogs to some of our inner
structures.  Although they do not report seeing conditions, the larger
pixel scale of their CCDs (0.46\arcsec/pixel) suggests their detection
efficiency could be lower than ours; they also noted that the
combination of seeing and their bulge subtraction method made reliable
detection of disks smaller than 5\arcsec{} in radius very difficult. 
Taken together, this suggests that the true inner-disk fraction for S0
galaxies may be even higher than we find.

However, Seifert \& Scorza's sample almost certainly includes some
barred galaxies; a number of their galaxies display the boxy or
peanut-shaped bulge isophotes now known to be the signature of a bar
seen edge-on \citep[e.g.,][]{combes90,kuijken95,bureau99,lutticke00}. 
Since we find that inner bars and disks are equally common in our
barred S0 galaxies, it is possible that some of Seifert \& Scorza's
``inner disks'' may in fact be inner bars.

In general, these previous studies appear to agree with our findings:
inner bars are rather common in early-type disk galaxies, occurring in
a quarter to a third of barred galaxies.  Inner disks are present with
a similar frequency --- but restricted almost exclusively to S0
galaxies.

\subsection{Implications for the Formation and Survival of Double Bars}

The random relative alignments of inner and outer bars
\citep[Section~\ref{sec:inner-bars};][]{buta93,w95} agrees with models
where inner and outer bars rotate at different speeds, as originally
argued by \citet{buta93} and \citet{friedli93b}.  The lack of
agreement between our observed inner bar + nuclear ring systems and
the predictions of \citet{shaw93} and \citet{heller96} is further
evidence against corotating secondary bars.  Thus, most double bars
are likely to be independently rotating systems.

The best-known scenario for forming independently rotating double bars is
essentially that of \citet{shlosman89}: a pre-existing large-scale bar
drives a gas inflow into the central kiloparsec of a galaxy; once
sufficient gas has accumulated, it becomes bar-unstable and a
dynamically decoupled (gaseous) secondary bar forms. 
\citet{friedli93b} and \citet{combes94} demonstrated this process with
n-body + hydrodynamical simulations; they showed that the stars in the
inner regions of the disk could participate in the secondary bar
instability, creating a secondary bar that was stellar as well as
gaseous.  Friedli \& Benz reported that the secondary bar's corotation
overlapped with the primary bar's ILR, as suggested by
\citet{pfenniger90}.  The correspondence between secondary bar sizes
and nuclear ring sizes noted in Section~\ref{sec:nr-db} agrees with
this scenario, especially if secondary bars end shortly before their
own corotation.

An alternate way of creating independently rotating double-bar systems
was suggested by \citet{friedli96b} and \citet{davies97}: formation of
two \textit{counter-rotating} bars in a galaxy with counter-rotating
stellar disks.  Under certain conditions --- specifically, with the
inner part of the stellar disk counter-rotating with respect to the
outer disk --- a double-bar system can form, with the inner bar
counter-rotating.  However, the high frequency of secondary bars
argues strongly against this model: \citet{kuijken96} found that the
fraction of S0 galaxies with significant stellar counter-rotation was
quite low (less than 10\%, and probably $\sim 1$\%).  Thus, most
double bars probably rotate independently and in the same direction.

The high frequency of secondary bars also suggests that secondary bars
are either long-lived, or that they (re)form frequently.  Because
\citet{friedli93b} found that their simulated secondary bars were
rather short-lived, dissolving after only about 1--2 rotations of the
outer bar ($\sim 250$--400 Myr), it has been argued that double bars
are transient \citep[e.g.,][]{knapen95}.  Friedli \& Martinet
suggested that the dissolution of the secondary bar was due to
bar-driven gas concentration in the center of the galaxy, in agreement
with an emerging theoretical picture of bar self-destruction, where
the growth of a sufficiently strong central mass concentration can
disrupt a bar \citep{hasan90,friedli93a,norman96}.  (In fact, in
Friedli \& Martinet's Model~III, \textit{both} bars dissolved at about
the same time.)  The conclusion from this line of reasoning, then, is
that we see a high double-bar frequency because secondary bars form
(and then reform after being destroyed) quite often, due to gas
inflows driven by primary bars.

However, there are problems with this scenario, both theoretically and
observationally.  First, \citet{combes94} found her inner bars lasted
rather longer than Friedli \& Martinet's (in some cases, more than 20
rotations of the outer bar).  Second, \citet{rautiainen99} reported
double-bar formation in n-body simulations, \textit{without any
dissipative component}.  Typically, the inner bar formed
\textit{first}, followed by the outer bar, and both were long-lived. 
(As pointed out in Section~\ref{sec:inner-bars}, the double-bar size
ratio in their simulations is much closer to that of our observed
double bars than the ratios of Friedli \& Martinet's simulations.) 
And third, recent high-resolution hydrodynamic simulations by
\citet{witold02} suggest that not all inner bars drive significant gas
inflows.  This casts doubt on the ability of inner bars to destroy
themselves as they build a central mass concentration.

If inner bars \textit{do} form as a result of gas inflows, and are
short-lived, then the prevalence of inner bars in S0 galaxies is
puzzling, given that S0 galaxies are relatively gas-poor.  Though some
of our inner bars appear to have significant gas inside, we do see
some ``clean'' inner bars with little or no dust
(Section~\ref{sec:inner-bars}).  The fact that inner bars are not
distinctly bluer than the surrounding bulge and outer bar suggests
that any star formation which might have accompanied (and consumed)
the bar-forming inner gas occurred at least a Gyr ago.  Finally, there
are two double-barred galaxies with substantial \textit{off-plane} gas
in the nuclear regions (NGC~3941 and NGC~7280), most likely the result
of accretion from other galaxies.  If the inner bars in these galaxies
were formed by (in-plane) gas inflow within the parent bars, then this
gas probably dissipated \textit{before} the interaction; otherwise,
the infalling gas would presumably have collided with the in-plane gas
and settled into the disk plane.  All of this implies that the
inner-bar systems we see are probably \textit{not} recently formed by
inflowing gas in the plane of the disk.

\section{Summary\label{sec:summary}}

\begin{enumerate}

\item In a survey of thirty-eight nearby S0--Sa barred galaxies in the
field, we found a total of ten double-barred systems.  We have
confirmed the existence of previously reported inner bars in NGC~2681,
NGC~2859, NGC~2950, and NGC~4314; NGC~2681 turns out to be
\textit{triply} barred.  We have also found six new double-barred
systems: NGC~718, NGC~2962, NGC~3941, NGC~3945, NGC~6654, and
NGC~7280.

\item Double bar systems are surprisingly common: at least a quarter,
and possibly as many as 40\%, of all barred S0--Sa galaxies harbor
secondary bars.  These bars show no particular preference for Hubble
type (within our narrow range) or primary bar strength.

\item Secondary bars are typically about 12\% the size of their
primary bars, with (deprojected) semi-major axes ranging from $\sim
250$ pc to 1 kpc (3--8\% of their galaxies' $R_{25}$).  The isophotal
ellipticities of secondary bars are lower than those of their parent
primary bars, which may be due to intrinsic roundness, or may be an
effect of being embedded within bright bulges.

\item Secondary bars probably rotate independently of, but in the same
direction as, their associated primary bars.  Their high frequency
even in gas-poor S0 galaxies, along with the absence of significant
in-plane dust in several secondary bars, suggests that they are
relatively long-lived structures.

\item We also find eight inner disks inside bars.  They form a
statistically distinct class in terms of their size relative to the
primary bar and their distribution among Hubble types; they cannot be
explained as secondary bars that happen to be aligned with the
galaxy's outer disk.  They range in (deprojected) radius from 6\% to
40\% of the size of the primary bars (260 pc to 1.7 kpc).  Inner disks
occur almost exclusively in S0 galaxies; none were found in Sa
galaxies.  Since secondary bars are about as common as inner disks in
S0 galaxies, it is possible that at least some of the inner disks seen
in edge-on galaxies \citep[e.g.,][]{seifert96} may actually be bars.

\item Gaseous nuclear rings, whether dusty or actively forming stars,
occur predominantly in Sa galaxies (24\% of the sample --- but 50\% of
the Sa galaxies --- have such nuclear rings).  We found three examples
of purely \textit{stellar} nuclear rings (in the S0 galaxies NGC~936,
NGC~2950, and NGC~3945), where the ring is made of stars similar in
color to the surrounding bulge and bar, and dust seems to be absent;
these may be the faded remnants of previous circumnuclear star
formation episodes.

\item The presence or absence of secondary bars appears to have no
significant effect on nuclear activity.  In contrast, nuclear spirals,
dusty or star-forming nuclear rings, and off-plane dust are very often
accompanied by LINER or Seyfert nuclei; this agrees with the findings
of \citet{martini99}.

\item Counting inner bars, inner disks, and stellar nuclear rings
together, at least 2/3 of all barred S0 (field) galaxies have some
central stellar structure inside the bar.  This may complicate
disk-bulge decompositions for such galaxies, and suggests that the
inner regions of early-type barred galaxies contain dynamically cool
and disklike components.

\end{enumerate}

\acknowledgments

We would like to thank Jay Gallagher for his help in formulating and
carrying out this observational program.  We also thank him and Witold
Maciejewski for numerous inspirational conversations and comments on
this work at various stages, and Marc Balcells and John Beckman for
comments on the final version.  Thanks also to Paul Martini and Isaac
Shlosman for stimulating conversations, and to the referee, Eric
Emsellem, for a very careful and constructive reading.  The work
reported here forms part of the Ph.D.\ thesis of Peter Erwin at the
University of Wisconsin--Madison.

This research has made extensive use of the NASA/IPAC Extragalactic
Database (NED) which is operated by the Jet Propulsion Laboratory,
California Institute of Technology, under contract with the National
Aeronautics and Space Administration.  We also made use of the
Lyon-Meudon Extragalactic Database (LEDA; http://leda.univ-lyon1.fr). 

Finally, this research was supported by NSF grants AST 9320403 and AST
9803114, and by grant AR-0798.01-96A from the Space Telescope Science
Institute, operated by the Association of Universities for Research in
Astronomy, Inc., under NASA contract NAS5-26555.


\clearpage

\onecolumn

\clearpage

\clearpage

\twocolumn

\clearpage
\onecolumn

\begin{deluxetable}{lll}
\tabletypesize{\scriptsize}
\tablewidth{0pt}
\tablecaption{Galaxies in the WIYN Sample\label{tab:galaxies}}
\tablecolumns{3}
\tablehead{
\colhead{Galaxy} & \colhead{Hubble Type (RC3)} & \colhead{Inner Structures}}
\startdata
NGC 718  & SAB(s)a                  & DB, NR\\
NGC 936  & SB(rs)$0^{+}$            & NR(s)\\
NGC 1022 & (R\arcmin)SB(s)a         & dusty\\
NGC 2273 & (R\arcmin)SB(s)a         & NS, NR\\
NGC 2655 & SAB(s)0/a                & dusty, OPG\\
NGC 2681 & (R\arcmin)SAB(rs)0/a     & DB, NR\\
NGC 2685 & (R)SB0$^{+}$ pec         & ID, OPG\\
NGC 2787 & SB(r)$0^{+}$             & dusty, ID, OPG\\
NGC 2859 & (R)SB(r)$0^{+}$          & DB, NR\\
NGC 2880 & SB$0^{-}$                & ID\\
NGC 2950 & SB(r)$0^{0}$             & DB, NR(s)\\
NGC 2962 & (R)SAB(rs)$0^{+}$        & DB\\
NGC 3032 & SAB(rs)$0^{0}$           & NS\\
NGC 3185 & (R)SB(r)a                & dusty\\
NGC 3412 & SB(s)$0^{0}$             & ID\\
NGC 3489 & SAB(rs)$0^{+}$           & dusty, NR\\
IC 676   & (R)SB(r)$0^{+}$          & dusty\\
NGC 3729 & SB(r)a pec               & dusty\\
NGC 3941 & SB(s)$0^{0}$             & DB, OPG\\
NGC 3945 & (R)SB(rs)0$^{+}$         & DB, NR(s), ID\\
NGC 4045 & SAB(r)a                  & dusty\\
NGC 4143 & SAB(s)$0^{0}$            & NS, ID\\
NGC 4203 & SAB$0^{-}$:              & OPG\\
NGC 4245 & SB(r)0/a:                & NS, NR\\
NGC 4310 & (R$\arcmin$)SAB$0^{+}$?  & dusty\\
NGC 4314 & SB(rs)a:                 & DB, NR, NS\\
NGC 4386 & SAB$0^{0}$:              & ID\\
NGC 4643 & SB(rs)0/a                & NS, ID\\
NGC 4665 & SB(s)0/a                 & \\
NGC 4691 & (R)SB(s)0/a pec          & dusty\\
NGC 5338 & SB0:                     & dusty\\
NGC 5377 & (R)SB(s)a                & NS, NR\\
NGC 5701 & (R)SB(rs)0/a             & NS\\
NGC 5750 & SB(r)0/a                 & dusty\\
NGC 6654 & (R$\arcmin$)SB(s)0/a     & DB\\
UGC 11920 & SB0/a                   & dusty\\
NGC 7280 & (R)SAB(r)$0^{+}$         & DB, OPG\\
NGC 7743 & (R)SB(s)$0^{+}$          & NS\\
\enddata
\tablecomments{Codes in the third column describe features found
within the (outer) bar for each galaxy: DB = galaxy is double barred
(i.e., inner/secondary bar found inside primary bar); ID = inner disk;
NR = nuclear ring; NR(s) = \textit{stellar} nuclear ring; NS = nuclear
spiral; OPG = evidence for off-plane gas or dust (such as a polar
ring); dusty = too dust-obscured to determine presence or absence of
central stellar structures.  (NGC~2787 has a large inner disk with
inclined dust lanes inside; the latter obscure the central regions.)}
\end{deluxetable}

\begin{deluxetable}{l||r|rrr|rr}
\tablewidth{0pt}
\tabletypesize{\footnotesize}
\tablecaption{Central Component Frequencies by Hubble Type and Bar Strength \label{tab:freq}}
\tablecolumns{7}
\tablehead{
\colhead{Component} & \colhead{All (38)} & \colhead{S0 (20)} &
\colhead{S0/a (10)} & \colhead{Sa (8)} & \colhead{SB (25)} &
\colhead{SAB (13)}}
\startdata
\cutinhead{Stellar Structures}
Inner Bar       & $26\pm7$\% (10) & $30\pm10$\% (6) & $20\pm13$\% (2) & $25\pm15$\% (2) &
$24\pm9$\%  (6) & $31\pm13$\% (4) \\
Inner Disk      & $21\pm7$\%  (8) & $35\pm11$\% (7) & $10\pm9$\% (1) & 
0\% (0)         & $24\pm8$\%  (6) & $15\pm10$\% (2) \\
No IB or ID     & $24\pm7$\%  (9) & $20\pm9$\%  (4) & $30\pm14$\% (3) & $25\pm15$\% (2) & 
$24\pm9$\%  (6) & $23\pm12$\% (3) \\
Nuc. Ring (s)   & $8\pm4$\%   (3) & $15\pm8$\%  (3) & 0\% (0) & 0\% (0) & 
$12\pm6$\% (3) & 0\% (0) \\
Dusty           & $34\pm8$\% (13) & $25\pm10$\% (5) & $40\pm15$\% (4) & $50\pm18$\% (4) & 
$36\pm10$\% (9) & $31\pm13$\% (4) \\

\cutinhead{Gaseous Structures}
Nuc. Ring (g)   & $21\pm7$\%  (8) & $10\pm7$\%  (2) & $20\pm13$\% (2) & $50\pm18$\% (4) & 
$20\pm8$\%  (5) & $23\pm12$\% (3) \\
Nuc. Spiral     & $24\pm7$\%  (9) & $15\pm8$\%  (3) & $30\pm14$\% (3) & $38\pm17$\% (3) & 
$28\pm9$\%  (7) & $15\pm10$\% (2) \\
Off-plane Gas   & $16\pm6$\%  (6) & $25\pm10$\% (5) & $10\pm9$\%  (1) & 0\% (0) &
$12\pm6$\% (3)  & $23\pm12$\% (3) \\
\cutinhead{``ILR'' features}
Nuc. Ring (all) & $29\pm7$\% (11) & $25\pm10$\% (5) & $20\pm13$\% (2) & $50\pm18$\% (4) &
$32\pm9$\% (8)  & $23\pm12$\% (3) \\
+ IB            & $39\pm8$\% (15) & $45\pm11$\% (9) & $20\pm13$\% (2) & $50\pm18$\% (4) &
$40\pm10$\% (10)  & $38\pm13$\% (5) \\
+ IB, ID        & $58\pm8$\% (22) & $75\pm10$\% (15) & $30\pm14$\% (3) & $50\pm18$\% (4) &
$60\pm10$\% (15)  & $54\pm14$\% (7) \\
+ IB, ID, NS    & $66\pm8$\% (25) & $85\pm8$\% (17) & $40\pm16$\% (4) & $50\pm18$\% (4) &
$68\pm9$\% (17)  & $62\pm13$\% (8) \\
\tablecomments{Percentages indicate what fraction of galaxies have a
particular type of central structure, as listed in Table~1.  These are
not exclusive categories, since some galaxies have multiple central
structures.  ``No IB or ID'' indicates the absence of an
inner/secondary bar or inner disk, though these galaxies may still
have nuclear rings, spirals, or off-plane gas.  ``Nuc.\ Ring (s)''
refers to stellar nuclear rings; ``Nuc.\ Ring (g)'' refers to gaseous
nuclear rings, where the ring is dusty and/or shows evidence of recent
star formation.  ``ILR'' refers to features which may indicate the
presence of an inner Lindblad resonance.  Error bars are
$\pm1$-$\sigma$, where $\sigma =$ binomial standard deviation. 
Numbers in parentheses are the number of galaxies in each category
(e.g., there are 20 S0 galaxies, 6 of which are double-barred).}
\enddata
\end{deluxetable}

\begin{deluxetable}{l|rrr}
\tablewidth{0pt}
\tabletypesize{\footnotesize}
\tablecaption{Nuclear Activity and Central Components\label{tab:agn}}
\tablecolumns{4}
\tablehead{
\colhead{Component} & \colhead{AGNs} &
\colhead{\hii{} Nuclei} & \colhead{Absorption-line Nuclei}}
\startdata
All Galaxies   & $61\pm9$\% (19)  & $19\pm7$\% (6)  & $19\pm7$\% (6) \\
\\
SB Galaxies    & $58\pm10$\% (14) & $17\pm8$\% (4)  & $25\pm9$\% (6) \\
SAB Galaxies   & $86\pm13$\% (6)  & $14\pm13$\% (1) & 0\% (0) \\
\\
Inner Bar      & $75\pm15$\% (6)  & 0\% (0)         & $25\pm15$\% (2) \\
Inner Disk     & $60\pm22$\% (3)  & 0\% (0)         & $40\pm22$\% (2) \\
None           & $56\pm17$\% (5)  & $22\pm14$\% (2) & $22\pm14$\% (2) \\
\\
ID or None     & $57\pm18$\% (8)  & $14\pm9$\% (2)  & $29\pm12$\% (4) \\
\\
Nuclear ring   & $82\pm12$\% (9)  & $9\pm9$\% (1)   & $9\pm9$\% (1) \\
\\
Nuclear ring (g) & $88\pm13$\% (7)  & $13\pm13$\% (1) & 0\% (0) \\
Nuclear spiral & $78\pm14$\% (7)  & $22\pm14$\% (2) & 0\% (0) \\
Off-plane gas  & 100\% (5)        & 0\% (0)         & 0\% (0) \\
NS or OPG      & $85\pm10$\% (11) & $15\pm10$\% (2) & 0\% (0) \\
NR(g), NS, OPG & $89\pm7$\% (17)  & $11\pm7$\% (2)  & 0\% (0) \\

\tablecomments{Distribution of nuclear activity for galaxies with
various types of inner structure, for the 31 galaxies with classified
nuclear spectra.  ``AGNs'' = Seyfert or LINER nuclei.  Percentages in
each row sum to 100\%, except for rounding; in parentheses are the
numbers of galaxies in each set (e.g., there are 24 SB galaxies with
nuclear classifications, and 14 of these --- 58\% --- are AGNs).  The
first set of three rows are exclusive categories, based on the
\textit{innermost} detected structure (so NGC 3945, which has an
inner bar and an inner disk, is only listed under inner bars). 
Error bars are 1-$\sigma$ binomial errors.}

\enddata
\end{deluxetable}

\end{document}